# *Pattern Matching for Self- Tuning of MapReduce Jobs*

This paper has been originally published as "On Using Pattern Matching Algorithms in MapReduce Applications" in IEEE International Symposium on Parallel and Distributed Processing with Applications, ISPA 2011, Busan, Korea, 26-28 May, 2011. We realized that the original title is not appropriate and cannot be found by people working in this area. Therefore, this text is just for changing the title; however, the original paper can be found at the rest of this text (starting from the next page). For citation, please cite the original title as:

*NB Rizvandi, J Taheri, AY Zomaya, "On Using Pattern Matching Algorithms in MapReduce Applications", IEEE International Symposium on Parallel and Distributed Processing with Applications, ISPA 2011, Busan, Korea, 26-28 May, 2011.*

# On Using Pattern Matching Algorithms in MapReduce Applications

Nikzad Babaii Rizvandi[1,2], Javid Taheri[1], Albert Y.Zomaya[1]

[1] Center for Distributed and High Performance Computing, School of IT, University of Sydney,

NSW 2006, Australia

[2] National ICT Australia (NICTA), Australian Technology Park,

NSW 1430, Australia

## ABSTRACT

In this paper, we study CPU utilization time patterns of several MapReduce applications. After extracting running patterns of several applications, they are saved in a reference database to be later used to tweak system parameters to efficiently execute unknown applications in future. To achieve this goal, CPU utilization patterns of new applications are compared with the already known ones in the reference database to find/predict their most probable execution patterns. Because of different patterns lengths, the Dynamic Time Warping (DTW) is utilized for such comparison; a correlation analysis is then applied to DTWs' outcomes to produce feasible similarity patterns. Three real applications (WordCount, Exim Mainlog parsing and Terasort) are used to evaluate our hypothesis in tweaking system parameters in executing similar applications. Results were very promising and showed effectiveness of our approach on pseudo-distributed MapReduce platforms

## 1. INTRODUCTION

Recently, businesses have started using MapReduce as a popular computation framework for processing large-scaled data in both public and private clouds. For example, many Internet endeavors are already deploying MapReduce platforms to analyze their core businesses by mining their produced data. Therefore, there is a significant benefit to application developers in understanding performance trade-offs in MapReduce-style computations in order to better utilize their computational resources [1].

MapReduce users typically run a few number of applications for a long time. For example, Facebook, which is based on Hadoop (Apache implementation of MapReduce in Java), is using MapReduce to read its daily produced log files and filter database information depending on the incoming queries. Such applications are repeated million times per day in Facebook. Another example is Yahoo where around 80-90% of their jobs is based on Hadoop [2]. The typical applications here are searching among large quantities of data, indexing the documents and returning appropriate information to incoming queries. Similar to Facebook, these applications are run million times per day for different purposes.

One of the major problems with direct influence on MapReduce performance is tweaking/tuning the effective configuration parameters [3] (e.g. number of mappers, number of reducers, input file size and so on) for efficient execution of an application. These optimal values not only are very hard to properly set, but also can significantly change from one application to another. Furthermore, obtaining these optimal values usually needs running an application for several times with different configuration parameters values; a very time consuming and costly procedure. Therefore, it becomes more important to find the optimal values for these parameters before actual running of such application on MapReduce.

Our approach, in this work, is an attempt toward solving this problem by predicting CPU utilization pattern of new application based on the already known ones in a database. More specifically, we propose a two-phase approach to extract patterns and find similarity in CPU utilization of MapReduce applications. In the first phase, profiling, few applications are run with different sets of MapReduce configuration parameters for several times to collect their execution/utilization profiles in a Linux environment. Upon obtaining such information –the CPU utilization time series of these applications–, their measurements noise is removed using a six order Chebyshev filter. Then, these CPU utilization values are stored in a reference database to be later used in the second phase, matching. In the matching phase, a pattern matching algorithm is deployed to   find similarity between stored CPU utilization profiles and the new application.

To demonstrate our approach, section 2 highlights the related works in this area. Section 3 provides some theoretical background for pattern matching and DTW. Section 4 explains our approach in which pattern matching is used to predict behavior of unknown applications. Section 5 details our experimental setup to gauge efficiency of our approach. Discussion and analysis is presented in section 6, followed by conclusion in section 7.

## 2. RELATED WORKS

Early works on analysing/improving MapReduce performance started almost since 2009; such as an approach by Zaharia et al [4] that addressed problem of improving the performance of Hadoop for heterogeneous environments. Their approach was based on the critical assumption in Hadoop that works on homogeneous cluster nodes where tasks progress linearly. Hadoop utilizes these assumptions to efficiently schedule tasks and (re)execute the stragglers. This paper designs a new scheduling policy to overcome these assumptions. Besides their work, there are many other approaches to enhance or analysis the performance of different parts of MapReduce frameworks, particularly in scheduling [5], energy efficiency [1, 6-7] and workload optimization [8]. A statistics-driven workload modeling was introduced in [7] to effectively evaluate design decisions in scaling, configuration and scheduling. The framework in this work is utilized to make appropriate suggestions to improve the energy efficiency of MapReduce. A modeling method was proposed in [9] for finding the total execution time of a MapReduce application. It uses Kernel Canonical Correlation Analysis to obtain the correlation between the performance feature vectors extracted from MapReduce job logs, and map time, reduce time, and total execution time. These features are critical for establishing any scheduling decisions. Recent works in [9-10] reported a basic model for MapReduce computation utilizations. Here, at first, the map and reduce phases are modeled using dynamic linear programming independently; then, these phases are combined to build a global optimal strategy for MapReduce scheduling and resource allocation.

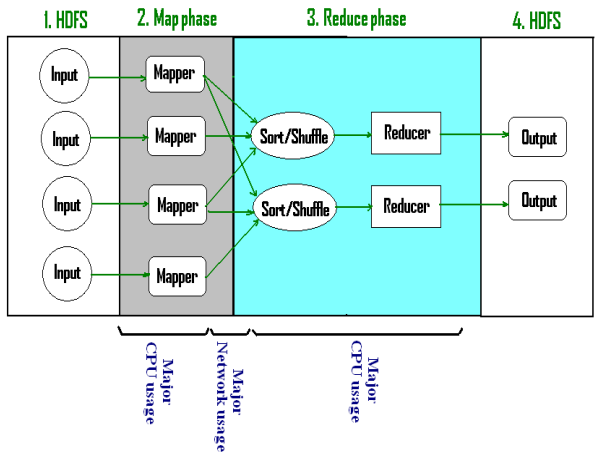

**Figure 1. MapReduce's flowchart**

The other part of our approach in this work is inspired by another discipline in which similarity of objects is also the center of attention and therefore very important – i.e., Speaker recognition. In speaker recognition (or signature verification) applications, it has been already validated that if two voices (or signatures) are significantly similar – based on a same set of parameters as well as their combinations –; then, they are most probably produced by a unique person [11]. Inspired by this well proved fact, our proposed technique in this paper hypothesizes the same logic with the idea of pattern feature extraction and matching, an area which is widely used in pattern recognition, sequence matching in bio-informatics and machine vision. Here, we extract the CPU utilization pattern of unknown/new MapReduce applications for a small amount of data (not the whole data) and compare its results with already known patterns in a reference database to find similarity. Such similarity will show how much an application is similar to another application. As a result, the optimal values of configuration parameters for unknown/new applications can be set based on the already calculated optimal values for known similar application in the database

## 3. THEORETICAL BACKGROUND

Pattern matching is a well-known approach – particularly in pattern recognition – to transform a time series pattern into a mathematical space. Such transformation is essential to extract the most suitable running features of an application before comparing it with reference application in a database to find its similar pairs. Such approaches have two general phases: (1) profiling phase, and (2) matching phase. In the profiling phase, the time series patterns of several applications are extracted. After applying some mathematical operations on these patterns, they are stored in a database as references during the matching phase. In matching phase, the same procedure is repeated for an unknown/new application first; and then, the time series of this application are compared with the saved time series of applications in database by using a pattern matching algorithm to find the most similar ones.

## 3.1 Pattern matching

One of the important problems in data mining is to measure similarity between two data series. Similarity measurement algorithms have been frequently used in pattern matching, classification and sequence alignment in bio-informatics. The measurement of similarity between two time series means to find a function: $SIM(X,Y)$ where $X$ and $Y$ are two time series with the with/without the same length. This function is typically designed as $0 \leq SIM(X,Y) \leq 1$, where greater values means higher similarities. In this case, $SIM(X,Y) = 1$ should be obtained for identical series only, whereas, $SIM(X,Y) = 0$ should reflect no similarity at all. Toward this end, similarity between two series is represented by defining a specific distance between them called the "similarity distance". Dynamic time warping (DTW) is one of the well-known and also efficient algorithm to calculate such similarity. This algorithm calculates the similarity between two different-length series by shifting and stretching the amplitude of both series and binding their time axes. The result is a path with minimum distance between two time series. To represent the similarity in percentage between these series, we calculate correlation coefficient on the outcome of DTW algorithm.

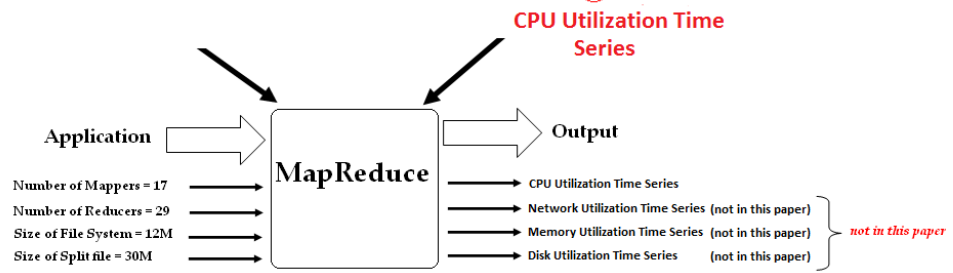

**Figure 2. The procedure of capturing CPU Utilization Time Series of a Mapreduce application**

### 3.1.1 Filtering and Magnitude Normalization

The captured CPU utilization time series (pattern) are usually noisy due to temporal changes coming from unknown devices states. Therefore, to increase accuracy, it is almost necessary to reduce noise from patterns before applying DTW. Among filtering methods in literature, we chose low pass Chebyshev filter as it is also widely used in speaker matching to pre-process CPU utilization time series of MapReduce applications. Also, we normalized our time series so that their values are always bounded between 0 and 1.

### 3.1.2 Dynamic Time Warping (DTW)

DTW is generally used to calculate the similarity distance between an input series with a reference series with a different length. A simple method to overcome unevenness of the series is to resample one series to match the other before comparison. This method is however usually results in unacceptable outcomes as the time serieses usually do not logically align correctly. DTW uses a nonlinear search to overcome this problem and map corresponding samples to each other. As a result, $X(t_1)$ might be aligned with $Y(t_1)$, while $X(t_2 + 8)$ is aligned to $Y(t_2 + 3)$. DTW uses the following mathematic recursive formulas to obtain similarity between two CPU utilization time series $X = [x_1, x_2,...,x_N]$ and $Y = [y_1, y_2,...,y_M]$ where $N \geq M$:

$$D(i,j) = \min \begin{cases} D(i, j-1) \\ D(i-1, j) \\ D(i-1, j-1) \end{cases} + d(x_i, y_j) \quad (1)$$

where $d(x_i, y_j)$ is the Euclidean distance between corresponding points in both series as:

$$d(x_i, y_j) = \|CPU(x_i) - CPU(y_j)\| \quad (2)$$

Here, $CPU(x_i)$ the value of CPU utilization in point $x_i$ in $X$. Results of these formulation is the $D(X,Y)$ matrix in which each of its elements – $D(i,j)$ – reflects the minimum distance between $[X(x_1), Y(y_1)]$ to $[X(x_i), Y(y_j)]$. As a result, $D(N,M)$ would reflects the similarity distance between $X$ and $Y$. In this case, if $N \geq M$, then, $Y'$ can be made from $Y$ with the same length as $X$ so that $Y'(t)$ is aligned with $X(t)$; $Y'$ is always made from $Y$ by repeating some of its elements based on $D(X,Y)$.

### 3.1.3 Similarity measurement

After finding the minimum distance path between two time series by DTW, which results in forming a new series $Y'$, the similarity between the series $X$ and $Y'$ is measured by calculating the correlation coefficient between these time series as [12]:

$$CORR(X, Y') = \frac{1}{N} \sum_{i=1}^{N} (x_i - \mu_X)(Y'_i - \mu_{Y'}) \quad (3)$$

This coefficient shows how much two series are correlated or similar. $CORR(X, Y') = 0$ indicates no similarity, while $CORR(X, Y') = 1$ reflects a perfect match; for other values, the greater correlation value, the higher similarity. In our approach, $CORR(X, Y') \geq 0.9$ is assumed an acceptable match – this value is set empirically.

In distributed computing systems, MapReduce has been known as a large-scale data processing or CPU intensive job [3, 14-15]. It is also well known that CPU utilization is the most important part of running an application on MapReduce. Therefore, optimizing the amount of CPU an application needs becomes important for customers to hire enough CPU resources from cloud providers and for cloud providers to schedule incoming jobs properly.

In this paper, we will study the similarity between CPU utilization time series of an incoming application with the reference applications in a reference database for different sets of configuration parameter values. If the CPU utilization time series of an unknown/new application is found to be adequately similar to CPU utilization time series of another application in database for the same and almost all sets of configuration parameters values; then, it can be assumed that the CPU utilization behavior of both applications are the same for other sets of configuration parameters values as well. This fact can be used in two ways: firstly, if the optimal values of the configuration parameters are obtained for one application, these optimal values can also be used for other similar applications too; secondly, this approach allows us to properly categorize applications in several classes with the same CPU utilization behavioral patterns.

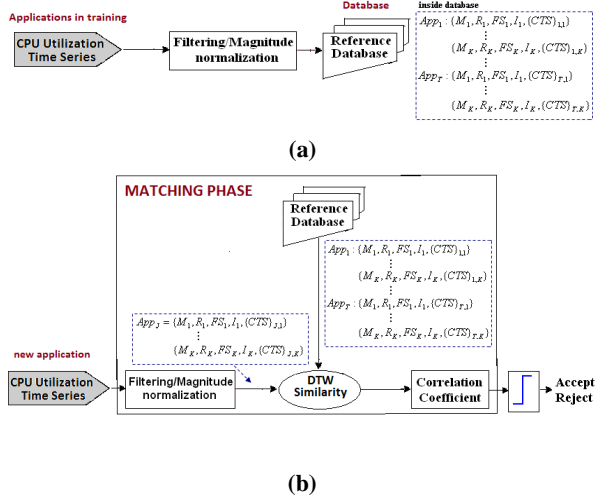

(a)

(b)

Figure 3. the block structure of profiling and matching phases used in our method.

## 4. PATTERN MATCHING IN MAPREDUCE APPLICATIONS

The goal in this paper is to propose an approach to predict CPU utilization pattern of an unknown/new application based on its similarity with already known ones in a reference database. Our approach is consisted of two phases: profiling and matching.

In the profiling phase, CPU utilization time series of several MapReduce applications in database is extracted. For each application, we generate a set of experiments with different values of four MapReduce configuration parameters on a given platform. These parameters are: number of mappers, number of reducers, size of split file systems and size of the input file. Within the system, we sample the CPU usage of each experiment from starting the map phase until finishing the reduce phase with time interval of one second; SysStat monitoring package in Linux is used to collect such CPU utilization time series (or patterns) [16]. Because of the temporarily changes, it is expected that CPU utilization time series of an experiment be affected by several noises. Therefore, a six-degree low pass Chebyshef filter is used to reduce such noises and smooth the time series. After reducing the noise, the time series with its related configuration parameters values are stored in reference database for future deployment (Figure 3-a). The algorithm related to the profiling phase has been shown in Figure 4-a. The algorithm indicates that for the first application, the application is run for the first set of configuration parameters on a small set of data. Then its CPU utilization Time Series (CTS) is captured by SysStat package. This application is then re-run for the second set of configuration parameters values and its CTS is also captured. This procedure is continued for all applications in database to profile different applications with several sets of configuration parameters values.

In the matching phase, the profiling procedure for gathering time series and de-noising of an unknown/new application is repeated and then followed by the several steps to find its similarity with already known application. These steps as shown in Figure 3-b calculate the pattern matching/similarity, and choosing the most suitable application. In calculating the pattern matching, after de-noising the CPU utilization time series pattern of the new application, DTW is applied – as pattern matching algorithm – to compare its time series pattern with those in the reference database. Eqn. 3 is then used to choose the most similar application. Here, it is assumed that CPU utilization pattern of the new application is most similar to the applications pattern with the highest similarity. Figure 4-b shows more details of this phase.

## 5. EXPERIMENTAL RESULTS

Three real applications are used to evaluate the effectiveness of our method; the first two are used to build the reference database, and only the last one is used for evaluation. Our method has been implemented and evaluated on a pseudo-distributed MapReduce framework. In such framework, all five Hadoop daemons (namenode, jobtracker, secondary namenode, datanode and task tracker) are distributed over cores/processors of a single laptop PC. Hadoop writes all files to the Hadoop Distributed File System (HDFS), and all services and daemons communicate over local TCP sockets for inter-process communication. In our evaluation, the system runs Hadoop version 0.20.2 that is Apache implementation of Mapreduce developed in Java [2]; the SysStat package is also concurrently executed in another terminal to monitor the CPU utilization time series of applications (in the native system) [16]. For an experiment with a specific set of configuration values (execution parameters), statistics are gathered from "running job" stage to the "Job complete" stage (arrows in Figure 3-left) with sampling time interval of one second. All CPU usages samples are then combined to form CPU utilization time series of an experiment.

Our benchmark applications are WordCount, TeraSort, and Exim mainlog parsing.

- *WordCount [17-18]:* This application reads data from a text file and, counts the frequency of each word. Results are written in another text file; each line of the output file contains a word and the number of its occurrence, separated by a TAB. In running a WordCount application on MapReduce, each mapper picks a line as input and breaks it into words. s. $<key, value>$ Then it assigns a $<key, value>$ pair to each word as $<word, 1>$. In the reduce stage, each reducer counts the values of pairs with the same $key$ and returns occurrence frequency (the number occurrence) for each word..

- *TeraSort :* This application is a standard map/reduce sorting algorithm – except for a custom practitioner that uses a sorted list of $N-1$ sampled $keys$ with predefined ranges for each reducer. In particular, all $keys$ with $sample[i-1] \leq key \leq sample[i]$ are sent to reducer i. This guarantees that the output of the $i^{th}$ reduce are always less than outputs of the $(i+1)^{th}$ reducer.

- *Exim mainlog parsing [19]*: Exim is a message transfer agent (MTA) for logging information of sent/received emails on Unix systems. This information that is saved in exim_mainlog files usually results in producing extremely large files in mail servers. To orginaize such massive amount of information, a MapReduce application is used to parse the data – in a exim_mainlog file – into individual transactions; each separated and arranged by a unique transaction ID..

We have tested our experiments on a Dell Latitude E4300 laptop with two processors: Intel Centrino model 2.26GHz, 64-bit; 2 x 2GB memory; 80GB Disk. For each application in both profiling and matching phases there are 50 sets of configuration parameters values where the number of mappers and reducers are chosen between 1 to 40 and the size of file system and the size of input file vary between 1Mbyte to 50Mbyte and 10MB to 500MB, respectively. Each experiment is executed on a small amount of data for these sets of configuration parameters to form a CPU utilization time series related to these parameters. Table 1 and figure 5 show the similarity measurement between CPU utilization patterns of Exim Mainlog Parsing application and other applications (WordCount and TeraSort). As can be seen, the highest similarities are observed between Exim and WordCount patterns – marked as red diagonally in the table. The reason for could be because of the similar nature of WordCount and Exim applications where both are applied on words in input text files; whereas, TeraSort tries to sort a large set of non-sorted input data that is a completely different problem with a different solution. Furthermore, the fact that these two applications are more similar when executed with identical set of parameters implies that Exim and WordCount applications are probably very similar in their execution behavior. Therefore, it can be concluded that WordCount and Exim would probably have almost the same CPU utilization pattern for other configuration parameters as well.

Our future plan is to analysis all resources (CPU, Disk and Memory); this requires three time series to be extracted for one application on our pseudo-distributed platform. Therefore, to find the similarity between two applications, three time series of the first application should be compared with their related time series from the second application. The complexity of our future work will be significantly greater than our current work in this paper; we also expect extra complexity when applications are run on real cluster. Toward this end, for an N-node cluster, three time series will be extracted from each node of cluster. Therefore, $3N$ time series is obtained for an application. As a result, the similarity problem will turn into comparing $3N$ time series from one application to its corresponding time series from another application. Due to the quadratic time and space complexity of DTW [20], working on $3N$ dimensions is computationally very expensive. Our idea to solve this problem is to extract wavelet coefficients of a time series and use them instead of original time series; this will result in a much shorter series than the original series. If all time series are transformed to wavelet domain by $M$ coefficients; then, the problem of similarity between two applications changes to obtain similarity between corresponding $3N$ wavelet coefficients series, with length $M$, of two applications. As two new series have the same length, then simple distance calculation instead of DTW can be utilized to find the similarity. However, using wavelet coefficients need to solve some challenging problems such as choosing an appropriate wavelet family as well as its number of coefficients ($M$).

|  | Exim MainLog Parsing | | | | |
|---|---|---|---|---|---|
|  |  | M=11, R=6, FS=20M, I=30M | M=21, R=30, FS=10M, I=80M | M=32, R=21, FS=30M, I=80M | M=42, R=33, FS=20M, I=60M |
| WordCount | M=11, R=6, FS=20M, I=30M | %94.3501 | %81.5059 | %81.8987 | %72.2611 |
|  | M=21, R=30, FS=10M, I=80M | %75.7111 | %92.1772 | %80.6990 | %88.3537 |
|  | M=32, R=21, FS=30M, I=80M | %81.7015 | %81.8771 | %91.8407 | %89.0751 |
|  | M=42, R=33, FS=20M, I=60M | %62.4942 | %83.8788 | %87.2693 | %93.9085 |
| Terasort | M=11, R=6, FS=20M, I=30M | %80.1223 | %80.8096 | %82.2029 | %69.5111 |
|  | M=21, R=30, FS=10M, I=80M | %58.4415 | %85.6312 | %77.3999 | %80.8949 |
|  | M=32, R=21, FS=30M, I=80M | %63.9299 | %77.0619 | %86.0416 | %86.7492 |
|  | M=42, R=33, FS=20M, I=60M | %89.4424 | %85.1708 | %82.1010 | %87.4424 |

TABLE 1. the result of similarity between Exim_manlog application as new application and WordCount and TeraSort as Reference applications for different sets of configuration parameters values

## 6. CONCLUSION
This paper presents a new approach that uses Dynamic Time Warping (DTW) to find the similarity among CPU utilization patterns of applications on MapReduce clusters. . After finding the minimum path between two patterns by DTW, the similarity is measured by calculating correlation coefficient between the patterns. Our experiments on three applications (WordCount, Exim Mainlog parsing and TeraSort) show that Exim Mainlog and WordCount have more similarity than Exim Mainlog and TeraSort for all sets of Mapreduce configuration parameters values.

## 7. ACKNOWLEDGMENT
The work reported in this paper is in part supported by National ICT Australia (NICTA). Professor A.Y. Zomaya's work is supported by an Australian Research Council Grant LP0884070.

**Profiling phase**
1. For $i^{th}$ application in database:
2.     For $j^{th}$ set of configuration parameters values ($M_j$, $R_j$, $FS_j$, $I_j$):
3.         Run application with the set of parameters on a small set of data
4.         Capture CPU utilization Time Series with Sysstat ($\{CTS\}_{i,j}$)
5.         De-noising the time series (Pre-processing)
6.         Add [$App_{i,j}$, $\{M_j, R_j, FS_j, I_j, \{CTS\}_{i,j}\}$] to database
7.     End
8. End

**(a)**

**Matching phase**
1. Take incoming new application
2. For $j^{th}$ set of configuration parameters values ($M_j$, $R_j$, $FS_j$, $I_j$):
3.     Run application with the set of parameters on a small set of data
4.     Capture CPU utilization Time Series with Sysstat ($\{CTS\}_{new,j}$)
5.     De-noising the time series (Pre-processing)
6. end
7. For $j^{th}$ set of configuration parameters values ($M_j$, $R_j$, $FS_j$, $I_j$):
8.     For $i^{th}$ application in database:
9.         Find matching between $\{CTS\}_{new,j}$ and $\{CTS\}_{i,j}$ using DTW
10.        Calculate correlation coefficient (CORR($\{CTS\}_{new,j}$,$\{CTS\}'_{i,j}$)
11.     End
12.     Pick the application with highest CORR if its CORR > 90%.
13. End
The application with the highest number of CORRs is the most similar application to the new application

**(b)**

**Figure 4. the detailed algorithms of profiling and matching phases.**

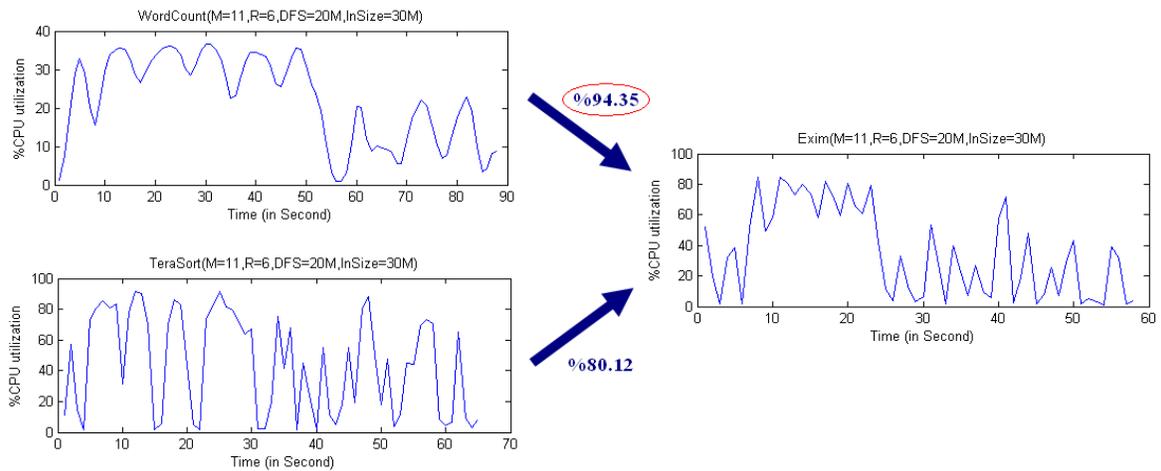

**(a)**

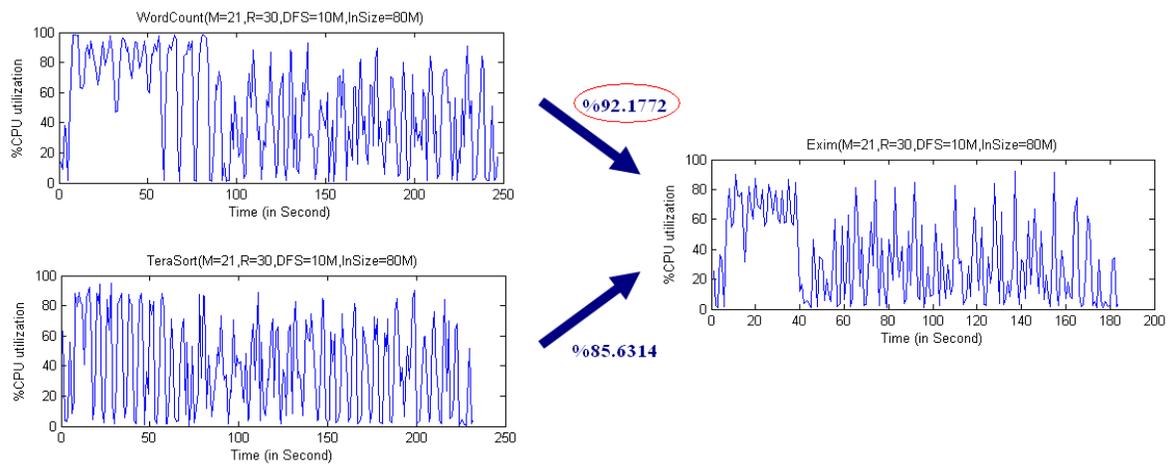

**(b)**

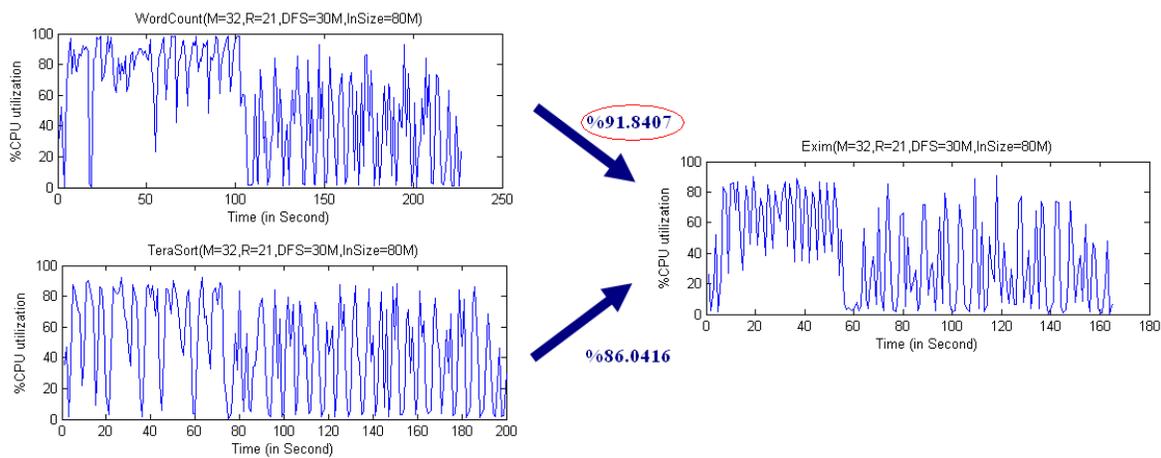

**(C)**

**Figure 6. Samples of similarity measurement between original Exaim-Manlog and WordCount and TeraSort time series in Reference database. As can be seen for the same configuration parameters values, Exim_mainlog and WordCount are more similar than Exim_mainlog and TeraSort for the same configuration parameters.**